\documentclass[%
 reprint,
superscriptaddress,
 amsmath,amssymb,
 aps,
pra,
]{revtex4-2}

\usepackage{graphicx}
\usepackage{dcolumn}
\usepackage{bm}

\begin{document}


\title{Self-bound clusters of one-dimensional fermionic mixtures}

\author{M. C. Gordillo}
\email{cgorbar@upo.es}
\affiliation{Departamento de Sistemas F\'{\i}sicos, Qu\'{\i}micos y Naturales, Universidad Pablo de
Olavide, Carretera de Utrera km 1, E-41013 Sevilla, Spain}
\affiliation{Instituto Carlos I de F\'{\i}sica Te\' orica y Computacional, Universidad de Granada, E-18071 Granada, Spain.}

\date{\today}

\begin{abstract}
Diffusion Monte Carlo calculations on the possibility of having self-bound one-dimen\-sional droplets of SU(6) $\times$ SU(2) ultracold fermionic mixtures are
presented.
We found that, even though arrangements with attractive interactions with only two spin types are not self-bound,  mixtures with at least three kinds of fermions
form stable small drops.  However,  that stabilization decreases for very tight confinements,  where a universal behaviour is found for Fermi-Fermi and Fermi-Boson clusters including attractive and  repulsive interactions. 
\end{abstract}

\maketitle

\section{Introduction}
In a liquid droplet the attractive interaction between particles should be balanced by repulsive forces in order to prevent the collapse of the system \cite{ferrier1}.  That is as true in a classical setting as when we are dealing with the ultradilute quantum drops \cite{frontiers2021} observed in Bose-Bose mixtures with either isotropic \cite{cabrera,soliton,freedrop} or dipolar interactions \cite{dipolar1,dipolar2}.  In the context of ultradilute cold atoms, that is the one we will be limited to in this work,  liquid is customarily understood as a synonym of self-bound, i.e, with a lower (more negative) energy than that of  the isolated units that conform it (see for instance Refs. \cite{ferrier1, frontiers2021, cabrera,njp2019,gregori1D,gregori1D2,giulia1}).  In particular, it does not imply any
particular kind of internal structure  different from that of a gas as it does in condensed matter physics.  The study of these self-bound systems started after a suggestion by Petrov \cite{petrov2015} (even tough there is at least a work \cite{bulgac} on the same topic that predates it),  whose theoretical study of binary bosonic arrangements with both attractive and repulsive interactions showed that terms of purely quantum origin prevented the collapse predicted by mean-field descriptions.  Examples of those liquid bosonic droplets and their stability limits could be found already in the literature \cite{frontiers2021,rep2020,gregori1D, gregori1D2,hui,giulia1,bruno}.

Beyond that frame, we can find works dealing with  mixtures of ultracold bosons and fermions
\cite{sal1,sal2,njp2016,scipost2019, njp2019,peacock},  whose stability for repulsively interacting bosons was found to be enhanced in one-dimensional (1D) setups \cite{njp2019}.  All those studies use in some form or another mean field approximations and are limited (but for the homogeneous arrangements in  Ref. \cite{peacock}) to the weakly interacting regime and to clusters in which the number of fermions is much smaller (ranging from an order of magnitude smaller to a single impurity) than the number of bosons. Considering all this, the goal of this work is to go further boson-boson and boson-fermion mixtures in two ways: first, we will explore the possibility of having stable self-bound  1D-drops made up exclusively of fermions with no bosons included.  In a sense,  this is a continuation of previous work on similar systems (see for instance  \cite{volosniev,ferro, matveeva,twoflavour,pilati}), but for a number of fermion species larger than two.  Second, since a set of distinguishable fermions can be considered effectively as a set of bosons, we will explore the stability limits of small 1D Bose-Fermi droplets in the strong interacting limit,  including correlations effects that are out of reach of mean field approximations.  We will also establish the minimum droplet composition to have self-bound droplets.

It is well known that a couple of spin-up and spin-down 1D-fermions that attract each other via a $\delta$ potential will pair to form a "molecule" irrespectively of the strength of their  interaction \cite{griffiths}.  This means that, in principle,  if we have a set of two different kinds of fermions with the same number of atoms each, we will have as many "molecules" as the number of particles of each set.  What we do not know is how those molecules interact with each other and how the internal spin composition of the 1D clusters
affects the possible phases we may have.
To study that, we are going to analyze mixtures of the fermionic isotopes of Ytterbium,  $^{173}$Yb and $^{171}$Yb \cite{pra77,cazalilla,taie}.  The atoms of the first species can have up to six different spin values (SU(6) symmetry),  while the second is a more conventional SU(2) arrangement \cite{cazalilla}, similar to $^6$Li.  To consider these systems have several advantages: first,  those mixtures have already been obtained \cite{taie}, what means that the conclusions of this work can be experimentally checked; second,  their masses, $m$, are close enough to be modeled by a single parameter.  This simplifies considerably the picture, since we do not have to take into account the effect of mass imbalance.
And last,   the strong attraction between the $^{173}$Yb and $^{171}$Yb atoms produces always molecules belonging to different isotopes, irrespective of the spin composition of the mixture.  This means than when we have spin-polarized $^{173}$Yb and $^{171}$Yb, the system is equivalent to a conventional SU(2) system made up of,  for instance, $^6$Li atoms that have only two spin states. 

Considering all of the above,  we have studied mixtures of $^{173}$Yb and $^{171}$Yb with different spin compositions 
in one-dimensional environments.
In accordance with previous literature,  and given the very low temperatures at which experiments involving ultracold atoms are done, we suppose those systems to be adequately described by the ground state (equivalent to consider T=0)
of the strictly 1D Hamiltonian \cite{su6su2,su6su2ol}:
\begin{eqnarray}\label{hamiltonian}
 H=\sum_{i=1}^{N_p}\frac{-\hbar^2}{2m}\nabla_i^2 + g_{1D}^{173-171} \sum_{i=1}^{N_{173}} \sum_{j=1}^{N_{171}} \delta(x^{173}_i - x^{171}_j)
 \nonumber \\
+ g_{1D}^{173-173} \sum_{b>a}^{6} \sum_{i=1}^{n_{173,a}} \sum_{j=1}^{n_{173,b}} \delta(x^{173}_{a,i} - x^{173}_{b,j})
 \nonumber \\
+ g_{1D}^{171-171} \sum_{b>a}^{2} \sum_{i=1}^{n_{171,a}} \sum_{j=1}^{n_{171,b}} \delta(x^{171}_{a,i} - x^{171}_{b,j}),
\end{eqnarray}
Here, $N_p$ is the total number of fermions, $N_{173}$ and $N_{171}$, are the total number of $^{173}$Yb and $^{171}$Yb atoms. 
In this work, we dealt only with balanced clusters, i.e., $N_{173}$= $N_{171}$ = $N_p$/2.
$n_{173,ab}$ and $n_{171,ab}$ are the number of atoms with spins $a$ and $b$.
The second term in Eq. \ref{hamiltonian} takes into account the interactions in which a member of a pair is a $^{173}$Yb atom and the other a  $^{171}$Yb particle, that do not depend on spin \cite{pra77},  so the position of the particles,  $x_{i,j}$,  are labeled by isotope type only.  On the other hand,  the next two terms deal with  $^{173}$Yb -$^{173}$Yb
and $^{173}$Yb -$^{171}$Yb pairs,  so the different spin types, $a$ and $b$ have to be included in the position  labels.
No interactions were considered for fermions of the same species, since those were kept apart by Pauli's exclusion principle.

Since we are interested in self-bound systems, we do not include a longitudinal confinement in the $x$ direction, something usually done by the introduction of a term of the type $1/2 m \omega_{\parallel}^2 x_i^2$.  The transverse confinement  that produces the 1D setup  is included in an effective way in the $g_{1D}$ parameters defined below \cite{Olshanii}, and it is customarily  described by a radial harmonic oscillator with differences between consecutive energy levels given by $\hbar \omega_{\perp}$.
Those levels are not populated beyond the ground state since $\omega_{\parallel}$ = 0 $<< \omega_{\perp}$ \cite{PhysRevA.78.033607}.  This means that it is much favorable for the particles of the system to accommodate any possible repulsive effective interactions by spreading in the longitudinal direction than to be promoted to the next transverse mode.  In addition,
when we have more than two spin species
the interaction between molecules is attractive (see Results section below), what implies that to promote some of those units to the next transverse mode we will have provide the energy to overcome that attraction.

The values of the $g_{1D}$'s  can be obtained via $g_{1D}^{\alpha,\beta} = -2 \hbar^2/m a_{1D}(\alpha,\beta)$, where the 1D-scattering lengths, $a_{1D}$,  are defined by \cite{Olshanii}:
\begin{equation} \label{a1D}
a_{1D}(\alpha,\beta) = -\frac{\sigma_{\perp}^2}{a_{3D}(\alpha,\beta)} \left( 1 - A \frac{a_{3D}(\alpha,\beta)}{\sigma_{\perp}} \right),
\end{equation}
with A=1.0326. $\sigma_{\perp} = \sqrt{\hbar/m \omega_{\perp}}$ is the  oscillator length  in the transversal direction, depending on the transversal confinement frequency  $\omega_{\perp}$.  This implies that the transverse confinement is harmonical and not of any other type, for instance box-like.  $a_{3D}(\alpha,\beta)$ stands for the three-dimensional set of scattering lenghts taken from Ref.  \cite{pra77}, i.e., 10.55 nm ($^{173}$Yb-$^{173}$Yb),
-0.15 nm ($^{171}$Yb-$^{171}$Yb) and -30.6 nm ($^{171}$Yb-$^{173}$Yb) \cite{pra77},  where the minus signs mean attractive interactions.  We considered that the only source of change for $a_{1D}$ comes from variations in the transverse confinement, since modifying the three-dimensional scattering lengths in Yb isotopes is problematic due to their particular electronic structure  \cite{cazalilla}. This means that the type of interaction between atoms is fixed by the sign of the scattering length in three dimensions: a negative $a_{3D}(\alpha,\beta)$ implies $g_{1D}^{\alpha,\beta} <$ 0 (atttactive) and vice versa.
                 
\section{Method}

To check if one can have self-bound 1D-drops of $^{173}$Yb and $^{171}$Yb atoms we  have to solve the Schr\"odinger equation derived from the Hamiltonian of Eq. \ref{hamiltonian} with free boundary conditions.  This means that no spurious periodicity and no confining external potential (such as one or several hard walls or an harmonic term) is imposed of the system.
Under such conditions, the solutions corresponding to that Hamiltonian in the absence of interaction between atoms are waves with any possible (positive) energy value in a continuous spectrum \cite{griffiths}.
To solve the full problem, we used the Fixed-node diffusion Monte Carlo method (FN-DMC)  \cite{hammond},  that provides us with an exact solution
within some statistical uncertainties, for the ground state (T=0) of a system of interacting fermions when the positions of the nodes of the exact wavefunction describing the system are known.
Fortunately,  in stricty 1D-systems we can have nodes only when two particles are exactly on top of each other \cite{ceperley}.
This information is easily included in the so called {\em trial} function, that is the initial approximation to the many-body real wavefunction the DMC algorithm needs.  The use of a Monte Carlo method allows us to go beyond mean-field approximations by introducing correlations betwen particles,  something that can be necessary to describe accurately dilute gas systems (see for instance Ref. \cite{petrov2015} and Ref. \cite{dy}, that provides a comparison of mean-field, quantum Monte Carlo, and  experiment for one of those systems).

Following Refs. \cite{su6su2,su6su2ol}, we used as a trial function:
\begin{eqnarray} \label{defa}
\Phi(x_1,\cdots,x_{N_p}) =
 \mathcal{A}[\phi(r_{11'}) \phi(r_{22'}) \cdots \phi(r_{N_{173},N_{171})}]
 \nonumber \\
\prod_{b>a}^{6}
\prod_{i=1}^{n_{173,a}} \prod_{j=1}^{n_{173,b}}  \frac{\psi (x_{a,i}^{173}-x_{b,j}^{173})}{(x_{a,i}^{173}-x_{b,j}^{173})} \nonumber \\
\prod_{b>a}^{2}
\prod_{i=1}^{n_{171,a}} \prod_{j=1}^{n_{171,b}}  \frac{\psi (x_{a,i}^{171}-x_{b,j}^{171})}{(x_{a,i}^{171}-x_{b,j}^{171})},
\end{eqnarray}
where $\mathcal{A}[\phi(r_{11'}) \phi(r_{22'}) \cdots \phi(r_{N_{173},N_{171})}] $ is the determinant of a square matrix
whose dimension is $N_{173} \times N_{171}$ \cite{pandaripande} (for a balanced cluster $N_{173}$ =  $N_{171}$)  and takes care of the interactions between pairs of particles of different isotopes.  $\phi(r_{ij'})$'s  are functions  that depend on the distance
between those particles $r_{ij'} = |x_i^{173} - x_{j'}^{171}|$ and are chosen as the  exact solution of the Sch\"odinger equation corresponding to a pair of non-confined 1D-particles
interacting with an attractive delta potential, i.e., \cite{griffiths}:
\begin{equation} \label{solution1D}
\phi(|x_i^{173}-x_{j'}^{171}|) = \exp \left[-\frac{|g_{1D}^{173,171}|}{2} |x_i^{173}-x_{j'}^{171}| \right].
\end{equation}
Since we are dealing with strictly 1D systems,  we can take that as an exact solution for the two-body interaction term without the regularization needed in higher dimensions. All this means that we can write any row of  $\mathcal{A}[\phi(r_{11'}) \phi(r_{22'}) \cdots \phi(r_{N_{173},N_{171})}] $ as:
\begin{widetext}
\begin{equation}
\exp \left(-|g_{1D}^{173,171 } |r_{i1'}/2 \right),   \exp \left(-|g_{1D}^{173,171 } |r_{i2'}/2 \right),  \cdots,  \exp \left(-|g_{1D}^{173,171 }|r_{i,N_{171'}}/2 \right)
\end{equation}
\end{widetext}
were $i$ stands for a particular $^{173}$Yb and the second index varies to consider all the atoms belonging to the $^{171}$Yb isotope, irrespectively of their spin.
The use of that structure implies that
the trial function is antisymmetric with respect to the interchange of two atoms of the same isotope \cite{pandaripande}.  It also means that when two same-species atoms are on the same position the trial function is zero, as it should for a couple of identical fermions.

The above determinant should describe accurately a system in which all $^{173}$Yb  atoms are spin-polarized
and the same is true of  all the $^{171}$Yb ones.   However, if we have, for instance,  two sets of $^{173}$Yb particles with different
spins this is not so.  The reason is that the atoms of those two different sets are {\em distinguishable} particles and not bound by the Pauli exclusion principle.
If we still use  $\mathcal{A}[\phi(r_{11'}) \phi(r_{22'}) \cdots \phi(r_{N_{173},N_{171})}]$ we are going to have nodes for positions for which $^{173}$Yb atoms of different spins are on top of each other, something that is, in principle,  not true.

To correct that, we have to look at how the determinant is built.
We can write down two consecutive rows accounting for the interaction of two distinguishable atoms at coordinates $x_i$ and $x_j$ in the $^{173}$Yb subset with all the $^{171}$Yb  particles (at coordinates $x_{1'},x_{2'}, \cdots,x_{N_{171'}}$) as:
\begin{center}
$\left|
\begin{array}{cccc}
\exp \left(-|g_{1D}^{173,171}|r_{i1'}/2 \right) &  \cdots & \exp \left(-|g_{1D}^{173,171}|r_{i,N_{171'}}/2 \right) \\
\exp \left(-|g_{1D}^{173,171}|r_{j1'}/2 \right) &  \cdots & \exp \left(-|g_{1D}^{173,172}|r_{j,N_{171'}}/2 \right)
\end{array}
\right|$ \label{row}
\end{center}
when $x_i \rightarrow x_j$, we can write
\begin{eqnarray} \label{def1}
\phi(r_{ik'})=\exp \left(-|g_{1D}^{173,171}| |x_i-x_{k'}|/2 \right) \nonumber \\ 
=  \exp \left(-|g_{1D}^{173,171}| |x_j+\Delta-x_{k'}|/2 \right)
\end{eqnarray}
with  $\Delta = x_i-x_j \rightarrow 0$. Expanding to the first order in  $\Delta$, we have:
\begin{eqnarray}\label{expansion}
\phi(r_{ik'}) = \nonumber \\
 \phi(r_{jk'}) - |g_{1D}^{173,171}| \frac{\exp \left(-|g_{1D}^{173,171}|r_{jk'}/2 \right) (x_j-x_{k'}) \Delta} {2 r_{jk'}}
\end{eqnarray}
This means
,that we can write $\mathcal{A}[\phi(r_{11'}) \phi(r_{22'}) \cdots \phi(r_{N_{173},N_{171})}]$
as a sum of two determinants,  the first one with two equal rows (and hence null) and a second one including the correction given by the last term of the right-hand-side of  Eq. \ref{expansion}.  With that in mind, we can see that the origin of the spurious node at $x_i-x_j \rightarrow 0$ is the dependence of all the elements
of that determinant  row on $\Delta$. This can be corrected by dividing those elements by, in this case, $x_i-x_j$. This is completely equivalent to consider a factor $1/(x_i-x_j)$ in the trial function.
We can repeat this procedure for any pair of distinguishable atoms in the $^{173}$Yb and
$^{171}$Yb ensembles.  This is the origin of the terms $(x_{a,i}^{\alpha}-x_{b,j}^{\alpha})$ in the denominator of Eq \ref{defa} \cite{su6su2,su6su2ol}.

$\psi (x_{a,i}^{\alpha}-x_{b,j}^{\alpha})$ (${\alpha}$ = 173,171) is a Jastrow function that introduces the correlations
between pairs of particles of the same isotope belonging to different spin species $a,b$.
Particles of the same isotope with the same spin are assumed to interact via Pauli exclusion only.
For the $^{173}$Yb-$^{173}$Yb pair,  we have chosen, following the previous literature for a pair of
repulsively-interacting particles  \cite{gregoritesis}:
\begin{widetext}
\begin{equation}\label{delta}
\psi (x_{a,i}^{173}-x_{b,j}^{173}) =\left\{
\begin{array}{l l}
 \cos(k[|x_{a,i}^{173}-x_{b,j}^{173}|-R_m])& |x_{a,i}^{173}-x_{b,j}^{173}| < R_m\\
 1 & |x_{a,i}^{173}-x_{b,j}^{173}|\ge R_m\\
\end{array}
\right.
\end{equation}
\end{widetext}
where $k$ was obtained by solving:
\begin{equation}
k a_{1D}(173,173)\tan(k R_m)=1,
\end{equation}
for each value of  $a_{1D}(173,173)$ deduced from Eq. \ref{a1D}  for a given transverse confinement.  $\omega_{\perp}$ was taken in the range 2$\pi \times$ 0-100 kHz, in line with previous experimental values \cite{pagano}.
The value of $R_m$ was the output of a variational calculation.  When the pair of particles of the same isotope attract each other, as in the $^{171}$Yb-$^{171}$Yb case, the Jastrow has
the form of Eq.  \ref{solution1D} \cite{gregoritesis,su6su2},
but with a different value of the defining constant, $g_{1D}^{171,171}$.  

DMC being an statistical method,  one have to be careful to avoid all possible sources of error.  The first one comes from spurious correlations.  Those could arise when we perform a single simulation and average the results after every single DMC step.  To avoid that,  for each Monte Carlo history comprising 2.5 10$^5$ steps (after thermalization),  we used configurations separated by 100 Monte Carlo steps to obtain the averages.  This means 2500 values instead of 2.5 10$^5$. In addition,   all the energies and other observables given below are the result of averaging six different Monte Carlo histories and the error bars will correspond to the standard deviation of those six values.  This follows closely the procedure of Ref. \cite{su6su2}.
The second source of error could come from the election in the number of walkers, $N_w$.  Following Refs \cite{su6su2} and \cite{bonin}, we performed an study of the convergence of the energy (after de-correlation) as a function of that parameter and found that any number of walkers equal or larger than 1000 produced the same values.  For instance,  for a cluster comprising 12 particles,  with two sets of three $^{173}$Yb atoms with different spins and six spin-polarized $^{171}$Yb atoms,  we have that for a time step
$\Delta \tau$ = 6 $\times$ 10$^{-4}$ $(\hbar \omega_{\perp})^{-1}$ the energy values per particle for $\omega_{\perp}$/(2$\pi$) = 25 kHz (the experimental value of Ref \cite{pagano}) were
-0.246 $\pm$ 0.007 $\hbar \omega_{\perp}$ for $N_w$ = 1000,   -0.249 $\pm$ 0.007 $\hbar \omega_{\perp}$ for $N_w$ = 1500,  and
-0.248 $\pm$ 0.007 $\hbar \omega_{\perp}$ for $N_w$ = 2000, all within the error bars of each other.  This justifies the election of 1000 walkers for all the energy results given below.  This is the same number used in Ref. \cite{su6su2} for a similar system.

\begin{figure}
\begin{center}
\includegraphics[width=0.8\linewidth]{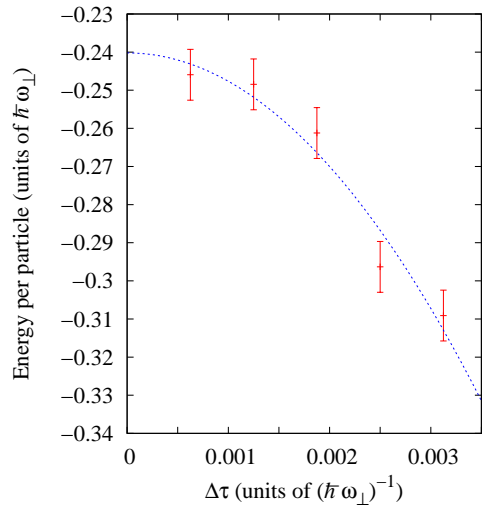}
\caption{Dependence of the DMC energy per particle on the DMC time step ($\Delta \tau$) for a cluster with two sets of three $^{173}$Yb atoms and a spin polarized subcluster with six $^{171}$Yb particles for $\omega_{\perp}$/(2$\pi$) = 25 kHz.  The number of walkers was $N_w$ = 1000.
}
\label{figa}
\end{center}
\end{figure}

The last possible source of error is the DMC time step,  $\Delta \tau$.  Fig. \ref{figa} shows the results of the extrapolation of the energy to the limit $\Delta \tau \rightarrow$ 0 for the same cluster considered above.  In that figure,  the dotted line is a least-squares fit to a quadratic form corresponding to the  propagator used \cite{boro94}.  The energy per particle, including the error bars derived from the fitting procedure is -0.240 $\pm$ 0.005 $\hbar \omega_{\perp}$, within the error bar of the value obtained for
$\Delta \tau$ = 6 $\times$ 10$^{-4}$  $(\hbar \omega_{\perp})^{-1}$.  From that and after having done similar studies in other clusters, we concluded that an adequate value for the DMC time step was $\Delta \tau$ =  6 $\times$ 10 $^{-4}$ $(\hbar \omega_{\perp})^{-1}$.  We have also tested that this
time step was large enough to provide a proper sampling of the all possible particle configurations.

\section{Results}

\begin{figure}
\begin{center}
\includegraphics[width=0.8\linewidth]{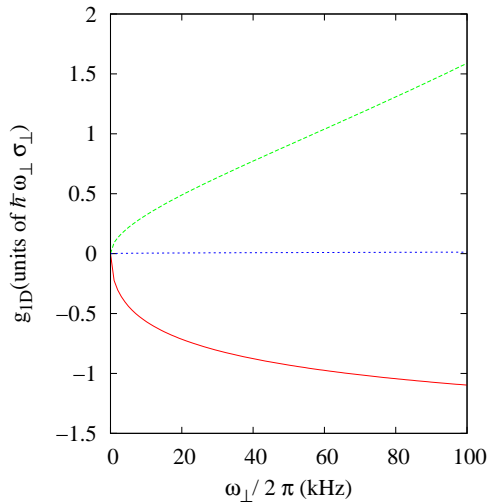}
\caption{Dependence of the $g_{1D}$ parameters of Eq. \ref{hamiltonian} on the transverse confinement,  $\hbar \omega_{\perp}$.  Full line,  $g_{1D}^{173-173}$; dashed line  $g^{1D}_{173-171}$; dotted line $g_{1D}^{171-171}$
}
\label{fig0}
\end{center}
\end{figure}

To solve the Schr\"odinger equation derived from the Hamiltonian in Eq. \ref{hamiltonian}, we need to deduce the $g_{1D}$ parameters for the interactions between Ytterbium isotopes.  That can be done with the help of Eq. \ref{a1D}, that relates the
physical magnitudes that define the system ($m$,  $\omega_{\perp}$, and the different three-dimensional
scattering lengths, $a_{3D}$),  to the corresponding $g_{1D}$'s.  Of those magnitudes, $m$ and $a_{3D}$ are fixed,  but
we can change  $\omega_{\perp}$ (0-100 2$\pi \times$ kHz) to modify the effective interaction between particles.  The result of such variation on the values of the $g_{1D}$'s
are displayed in Fig. \ref{fig0}.
There,  we can see that the $^{173}$Yb-$^{173}$Yb
interaction is always repulsive, while the $^{173}$Yb-$^{171}$Yb is always attractive and of the same order of magnitude.
At the same time,  $g_{1D}^{171,171} \sim$ 0 in all the $\omega_{\perp}$  range.  It is also important to stress that we have chosen to consider only interaction parameters compatible with experimental conditions. This means that, for instance,  the ratio $g_{1D}^{173,173}$/$g_{1D}^{173,171}$
cannot be varied arbitrarily,  being fixed by Eq. \ref{a1D}.  
With that in mind, we solved the Schr\"odinger equation corresponding to the Hamiltonian in Eq. \ref{hamiltonian} for a set of balanced clusters with different number of total particles, $N_p$ =4,6,8,10,12.

We started with arrangements in which all the
$^{173}$Yb and  $^{171}$Yb atoms are spin-polarized, i.e., belong to the same spin species. That system would be similar to a set of paired spin-up and spin-down $^6$Li atoms.   For {\em all} values of the transverse confinement, the total energy was -$N_p E_b$/2, with $E_b$ the binding energy between
$^{173}$Yb-$^{171}$Yb pairs. That energy is  \cite{griffiths,PhysRevA.78.033607}:
\begin{equation}
\label{eb}
E_b =
\frac{\hbar \omega_{\perp}}{4} \left(\frac{g_{1D}^{173,172}} {\hbar \omega_{\perp} \sigma_{\perp}} \right)^2
\end{equation}
This means that the particles arranged themselves in pairs
formed by atoms of different isotopes with no attraction between them.  This is similar to what happens in three-dimensions  but with a subtle difference: for instance, in Ref. \cite{jordibec},  the system had periodic boundary conditions, and the total energy, after substracting the corresponding to the binding of the pairs, was positive, not zero as in our free boundary conditions arrangement.
Those molecules do not interact with each other due to the avoidance of {\em both} elements of the pair by fermions of the same species in other pairs brought about by Pauli exclusion principle.
That precludes the formation of  self-bound drops, and it should be a common property of any SU(2) balanced system with very short-range  ($\delta$ or its counterpart in higher dimensions) attractive interactions.  Thus,  to check if a droplet is stable we will have to verify that the total  energy of the cluster is below -$N_p E_b$/2, i.e. more negative than the corresponding to a set of non-interacting paired molecules. This is exactly the same criterion used in $^3$He drops \cite{pandaripandehe3,navarro,sola}. 

\begin{figure}
\begin{center}
\includegraphics[width=0.8\linewidth]{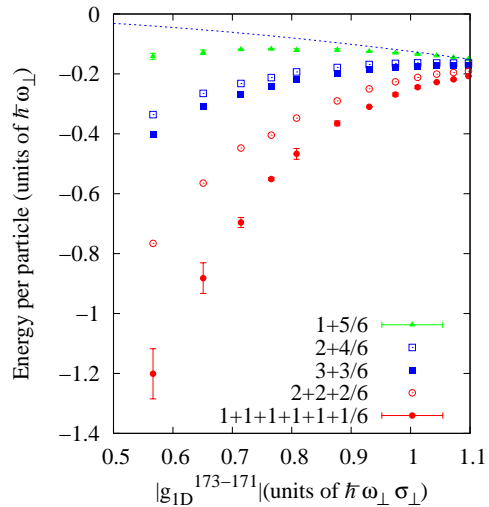}
\caption{Energy per particle in units of $\hbar \omega_{\perp}$ for a set of balanced clusters with a total number of particles, $N_p$,  equal to 12 as a function of the $|g_{1D}^{173-171}|$ parameter.  The clusters are named as
$\sum_{a=1}^{s_{173}} n_{173,a} / \sum_{b=1}^{s_{171}} n_{171,b}$.  The energies of two of the  droplets were given with their corresponding error bars. When not shown, those bars were of the same size that the ones displayed for the same values of $|g_{1D}^{173-171}|$.  The dotted line corresponds to $E_b$/2 = - $(g_{1D}^{173-171})^2$/8.
}
\label{fig1}
\end{center}
\end{figure}

Fig. \ref{fig1} reflects the change in the previous  situation when we have more than one spin species for the $^{173}$Yb set of atoms and keep the $^{171}$Yb spin-polarized.
There, we display the energy per particle for balanced clusters with $N_p$ = 12 in units of $\hbar \omega_{\perp}$ as a function of the absolute value of the $g_{1D}^{173-171}$ (in units of $\hbar \omega_{\perp} \sigma_{\perp}$).  Those units are the ones customarily used in the literature,  and can be translated into experimental physical magnitudes via Eq. \ref{a1D}.  From Eq. \ref{eb},  we can deduce that the energy per particle of a spin-polarized cluster should be  -$E_b$/2 = -$(g_{1D}^{173,171})^2$/8, value displayed in Fig. \ref{fig1} as a dotted line.
What we see is that for small values of  $|g_{1D}^{173-171}|$, corresponding to relatively loose confinements, the energy per particle for any cluster composition becomes appreciably more negative than the one for the case of two spin-polarized isotopes,  but that for tighter confinements (larger $|g_{1D}^{173-171}||$'s) gets progressively  closer to that number.

From the analysis of Fig. \ref{fig1} we can obtain several conclusions.  First, we can see that it is enough to flip a single spin in the $^{173}$Yb set to obtain a self-bound droplet.  This can be deduced from the result for a 1+5/6 cluster,  but it is also applicable to
an $^{171}$Yb  flip in a 6/5+1 arrangement,  whose energies are not given by simplicity.   This stabilization is due to the relaxation of the Pauli-related restrictions,  that allows, in the first case,  a single $^{171}$Yb atom to be close to several $^{173}$Yb atoms of different spins.   At the same time, those $^{173}$Yb atoms can be arbitrarily close together since they belong to different species.  The only price to pay will be a repulsive $^{173}$Yb-$^{173}$Yb  interaction (not avoidance as for undistinguishable fermions)  that can be counterbalanced by the attraction between atoms of different isotopes.
Obviously,  the effects of this relaxation increase with the number of different spin species, making the arrangements progressively more stable.   Thus,  the system with the lowest energy per particle comprises six {\em distinguishable} $^{173}$Yb and six spin-polarized $^{171}$Yb, and it is equivalent to a Fermi-Bose arrangement with a Fermi/Bose ratio 1:1. This system is stable for all the $g_{1D}^{173-171}$ (and hence $\omega_{\perp}$) values considered in this work.  The mean field approximation used
 in Ref. \cite{njp2019} would preclude the stability of those clusters for the lowest values of $g_{1D}^{173-171}$.

\begin{figure}
\begin{center}
\includegraphics[width=0.8\linewidth]{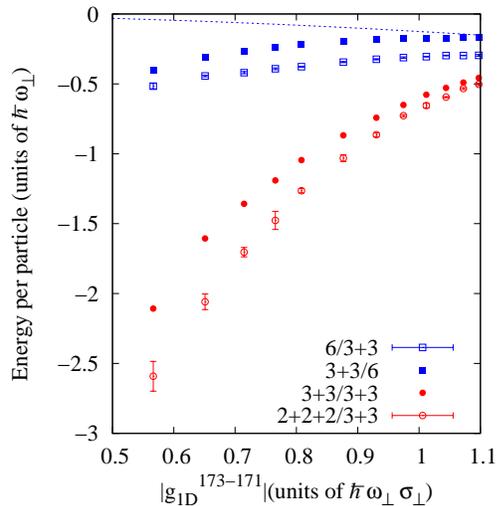}
\caption{Same as in the previous figure but for clusters of different compositions.  In the 6/3+3 cluster, the error bars are of the same size as the symbols.
}
\label{fig2}
\end{center}
\end{figure}

\begin{figure}
\begin{center}
\includegraphics[width=0.8\linewidth]{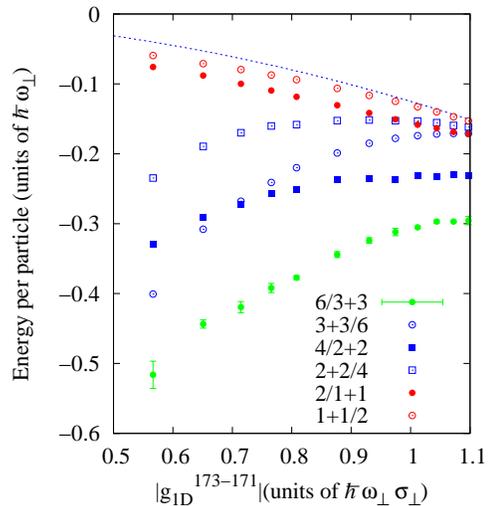}
\caption{Energy comparisons between clusters of similar compositions but different total number of particles.
}
\label{fig3}
\end{center}
\end{figure}

The conclusions of the previous paragraphs are fully supported by the study of similar or smaller clusters of different compositions.
For instance, in Fig. \ref{fig2} we can see the effect that the consideration of different number of species have  for $N_p$=12.  The 3+3/6 case is repeated from the previous figure to serve as a stick of comparison.  The overall behavior of the droplets is similar to the one shown in Fig.  \ref{fig1}: there is a sizeable stabilization for relatively loose transverse confinement, stabilization  reduced for very thin tubes.  We can see also that the larger the number of spin species is, the lower the energy per atom.  In addition,  clusters with the same number of species but in which we kept the $^{173}$Yb atoms spin-polarized are more stable than the ones with spin-polarized $^{171}$Yb's.  This is due to the weak attraction between $^{171}$Yb-$^{171}$Yb pairs that kicks in  when the Pauli restrictions are relaxed.  A similar set of rules can be applied to understand the clusters displayed in Fig. \ref{fig3}.  The only
additional information is the increasing of the stabilization of the cluster with size. There, we can see that even clusters with very small number of particles are stable.  This justifies us in the use of clusters with $N_p$ = 12 particles, since one would expect further increases in stabilization with size.

\begin{figure}
\begin{center}
\includegraphics[width=0.8\linewidth]{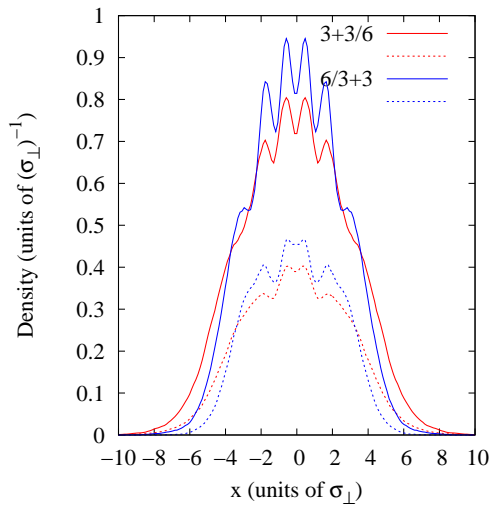}
\caption{Density profiles of the 3+3/6 and 6/3+3 clusters for $\omega_{\perp}$ = 2$\pi \times$ 25 kHz. Solid lines, spin-polarized isotopes; dotted lines, other species. The density profiles are normalized to the number of particles, i.e., six or three, respectively.  The center of the cluster corresponds to the position of its center of mass.
}
\label{fig4}
\end{center}
\end{figure}

\begin{figure}
\begin{center}
\includegraphics[width=0.8\linewidth]{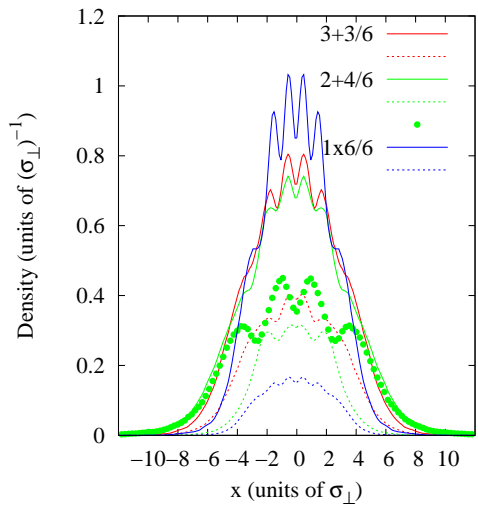}
\caption{Same as the previous figure but for clusters of different compositions.  Full lines, spin-polarized ytterbium; dotted lines, minority (or averages of equally distributed) $^{173}$Yb atoms; full circles,  majority component in the 2+4/6 clusters. All the densities are normalized to their respective number of atoms.  1x6/6 is short for 1+1+1+1+1+1/6. Error bars are of the size of the symbols and not displayed by simplicity.
}
\label{fig5}
\end{center}
\end{figure}

\begin{figure}
\begin{center}
\includegraphics[width=0.8\linewidth]{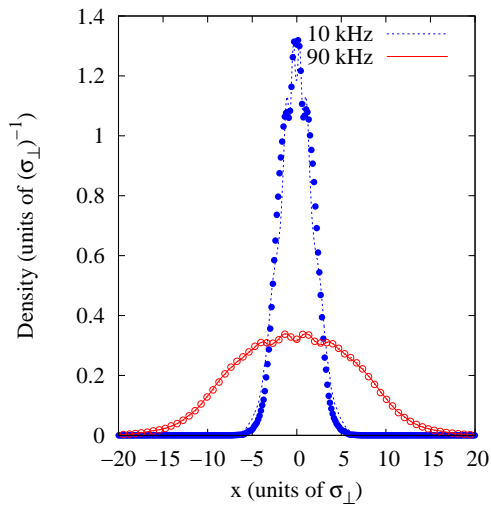}
\caption{ Density profiles for 3+3/6 clusters for two different values of the transverse confinement: $\omega_{\perp}$ = 2 $\pi \times$ 10 kHz (dotted lines and full spheres) and 2$\pi \times$ 90 kHz (full lines and open circles). Lines correspond to the $^{171}$Yb isotope and symbols to the $^{173}$Yb one.  In opposition to the case of Fig. \ref{fig4} we represented the total density of
the last type of  atoms, normalizing the profile to six.
}
\label{fig6}
\end{center}
\end{figure}

That stabilization does not imply the collapse of the clusters. In Fig. \ref{fig4} we can see the density profiles (number of perticles per unit length) corresponding to the 3+3/6 and 6/3+3 droplets.  Those were calculated from DMC configurations,  and could be appreciably different from the obtained by using a mean-field method  \cite{dy}.  Since free boundary conditions were used,  the particles were able to wander freely in 1D-space.  To avoid that effect,  we calculated the density profiles taking as the origin of coordinates the center of mass of the cluster.  The signature of a stable drop is then a finite width of those profiles,  something that does not happen for spin-polarized clusters of $^{173}$Yb and $^{171}$Yb.  According to that prescription,  all the clusters with energies smaller than the corresponding to a set of independent molecules have constant width, as can be been in Figs. \ref{fig4},\ref{fig5} and \ref{fig6}.  The form of all the profiles, with their maxima at the center of the cluster
implies that such small clusters are not made up of smaller subunits close together.
In Fig. \ref{fig4} the solid lines correspond to the spin-polarized isotope:  $^{171}$Yb in the first case, and $^{173}$Yb
in the second.  The profiles are normalized to the number of particles in that part of the arrangement, i.e., six.
On the other hand,  the dotted lines are the averages for the 3+3 part of those systems,  and their areas as half as much as those of their polarized counterparts.  The value of the transverse frequency was fixed to 2$\pi \times$25 kHz,  the same as the one in the experimental work of Ref. \cite{pagano} for a set of $^{173}$Yb atoms.  What we observe is that the width of the clusters is basically the same, but a little bit thinner in the 6/3+3 case, due to the attraction of the $^{171}$Yb-$^{171}$Yb pairs with different spins, but in any case different from zero.  The same can be said of the arrangements 
displayed in Fig.  \ref{fig5}. In this last case, the differences can be ascribed to the different number of spin species of the $^{173}$Yb isotope.  In all cases we have stable self-bound finite-size drops.

We can also see the influence of the transverse confinement in the shape of the density profiles. In Fig. \ref{fig6} we can see what happens to the 3+3/6 profile when squeezed in that direction.  We have chosen this particular arrangement because,  as can be seen in Fig. \ref{fig3}, the energy per particle
is noticeably  below -0.5$E_b$ for all the values of $|g_{1D}^{173,171}|$ considered.  At the same time, the variations in the shape of the profiles are fairly representative of what we can find in other cases.  In that figure, we represented both the $^{171}$Yb profiles (lines) and the sum of the $^{173}$Yb ones (symbols), i.e., this last profile is normalized to six instead of the three in Figs. \ref{fig4} and \ref{fig5}.  What we observe is that, while at low confinements, the total $^{173}$Yb and $^{171}$Yb distributions are different,  at $\omega_{\perp}$ = 2$\pi \times$90 kHz, both of them are basically identical.  This can be the product of a increasing in the repulsion among the atoms of the SU(6) isotope,  that makes the system more similar to a set of balanced $^6$Li atoms.   In any case, the similarity is not complete, since there is an energy excess that stabilizes the Ytterbium clusters, and will not do the same for a set of $^6$Li atoms.  This figure,  together with the previous ones can be used to attests the one-dimensionality of the system: the minimum spread on the longitudinal direction corresponds to $\sim$10 $\sigma_{\perp}$ for  $\omega_{\perp}$/(2$\pi$) = 10 kHz ,  with typical values of 15-20 $\sigma_{\perp}$ at the experimental frequency of $\omega_{\perp}$/(2$\pi$) = 25 kHz (see Figs. \ref{fig4}-\ref{fig5}) and going up to $\sim$ 40 $\sigma_{\perp}$ at $\omega_{\perp}$/(2$\pi$) = 90 kHz.  Those are larger that the value corresponding to the transverse width, by definition,  $\sim \sigma_{\perp}$.

\begin{figure}
\begin{center}
\includegraphics[width=0.8\linewidth]{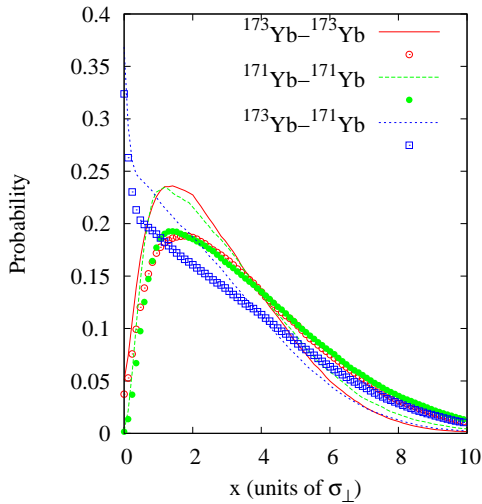}
\caption{Probability of finding another particle at a distance $x$ from the first one  for "1+1+1+1+1+1/6" (lines) and "3+3/6" (symbols)  clusters.   The profiles are normaized to one, including the tails beyond $x$ = 10$\sigma_{\perp}$, not shown for simplicity.
}
\label{fig7}
\end{center}
\end{figure}

Last, in Fig. \ref{fig7}, we show the probability of finding another particle at a distance $x$ of a given one.  This gives us information about the correlations between pairs.  Those probabilities were calculated  for "1+1+1+1+1+1/6" (lines) and "3+3/6" (symbols) clusters for a transverse confinement of 2$\pi \times$25 kHz.  Other arrangements are qualitatively similar and not shown for simplicity.  The main features of this observable are covered by representing the probabilities corresponding to all the possible isotope pairs, i.e., all atoms of the same isotope were lumped together.  Since in both clusters the $^{171}$Yb is spin-polarized,  the zero value of that function for a $^{171}$Yb-$^{171}$Yb pair for $x \rightarrow 0$ is simply a consequence of Pauli's exclusion principle.   On the other hand,  since at least part of the $^{173}$Yb atoms belong to different spin species, in that limit the probability of having a $^{173}$Yb-$^{173}$Yb pair is different from zero.  The position of the maxima in both $^{173}$Yb-$^{173}$Yb and $^{171}$Yb-$^{171}$Yb functions, roughly similar to each other,  reflects the typical distance between different molecules.
The existence of that molecules can be deduced from the maxima in the $^{173}$Yb-$^{171}$Yb probability function for $x \rightarrow 0$.

\section{Conclusions}

In this work, we have study the possibility of the existence of one-dimensional self-bound mixtures of Ytterbium fermionic isotopes.
We have chosen this system because those mixtures have already been produced \cite{taie} and the consideration of isotopes allowed us to use the same mass for all the atoms in the clusters, simplifying their description.  To be realistic,  experimentally-derived parameters were used to describe the interactions between different species.  Since the goal of this work was to establish the possibility of having self-bound clusters and the energy per particle was found to decrease when increasing the number of particles, we stop at a $N_p$ = 12,  since our results indicate that we can expect the trend to continue for larger $N_p$'s.  In particular,  the study of convergence (if any) to the thermodynamic limit ($N_p \rightarrow \infty$) is out of the scope of this work.  Sets of 1D-fermions of this size can be obtained experimentally \cite{smallcluster},  so the size of the clusters presented here is not an obstacle to their experimental realization.

Several interesting conclusions can be afforded by the analysis of the data above.  First, the existence of self-bound 1D-droplets made up of two spin-polarized sets of the same number of fermionic atoms with attractive $\delta$ interactions is not possible.  This is so because they form molecules that exclude other pairs in the vicinity due the double effect of the Pauli exclusion principle between atoms of the same species in different molecules.   Even though this is similar to what happens in three-dimensional SU(2) systems with short-range interactions  in periodic boundary conditions, \cite{jordibec}  this effect has not been previously described as such.   Secondly,  to flip the spin of a single atom is enough to produce a stable drop.  This can be seen with the help of Figs. \ref{fig1} and  \ref{fig3}, that allows to see that a 1+5/6, 2/1+1 and 1+1/2 clusters are stable.  In addition, we have verified that the 6/5+1, 4/3+1,3+1/4 clusters are also self-bound  and have compact density profiles.  Moreover,  a close inspection of Figs. \ref{fig1}-\ref{fig3} indicates that any cluster with at least three fermionic species is stable, providing that at least one of the $\delta$ interaction between species is attractive.  This is a general conclusion that could be experimentally tested.

The third relevant finding of this work has to do with the behavior of the clusters at very tight confinements, i.e, for large values of
$|g_{1D}^{173,171}|$ and $g_{1D}^{173,173}$ (see Fig. \ref{figa}).                                                                                                     
With the help of Figs. \ref{fig1}-\ref{fig3} we can see that the tighter the confinement, the closer the value of the energies per particle to -$E_b$/2 for $\sum_{a=1}^{s_{173}} n_{173,a} $/($N_p/2$) arrangements.  This means clusters with several spin values for $^{173}$Yb and spin-polarized in its $^{171}$Yb part. This suggests that the $\omega_{\perp} \rightarrow \infty$ energy limit for those clusters is universal and equal to -$E_b N_p$/2.  Since a set of $N_p$/2 distinguishable fermions is akin in this context to a set of bosons,
that limit would also apply to a 1D-Fermi-Boson mixture in which the boson-boson interaction is repulsive.  This is corroborated
by the behavior of the 1+1+1+1+1+1/6 and 1+1/2 arrangements, displayed in Figs. \ref{fig1} and \ref{fig3}.   In a sense, this is  equivalent to the Tonks-Girardeau limit for 1-D repulsively interacting single fermions \cite{Girardeau} but for pairs of molecules.  In that limit, there is no difference between the energies of a set of fermions or bosons for harmonically confined systems.  As in the Tonks-Girardeau gas,  for very tight confinements we are in the strong interaction limit, something that cannot be dealt with the mean-field approximations used for Fermi-Bose gases in the previous literature \cite{njp2019}.
However, the situation is slightly different for ($N_p$/2)/$\sum_{a=1}^{s_{171}} n_{171,a}$ clusters, in which fermions with different spins attract each other, even slightly.  Then, even tough there is still an energy limit for very tight confinemente, that limit  is  and lower 
than the corresponding to -$E_b$/2 per particle (see Fig. \ref{fig3}),  as a result of the residual attraction between molecules. 

Summarizing,  this study opens the door to consider new behaviors for attractively interacting fermionic mixtures beyond binary  compositions both in 1D and higher dimensions.  Those studies need not be limited to mixtures of Yb isotopes, but could be extended to systems made up of atoms with different masses providing we know all the experimentally relevant parameters
($m$'s, scattering lengths and transverse confinements).

\begin{acknowledgments}

We acknowledge financial support from Ministerio de
Ciencia e Innovación MCIN/AEI/10.13039/501100011033 
(Spain) under Grant No. PID2020-113565GB-C22
and from Junta de Andaluc\'{\i}a group PAIDI-205.  
We also acknowledge the use of the C3UPO computer facilities at the Universidad
Pablo de Olavide.
\end{acknowledgments}

\bibliography{droplets}

\end{document}